\documentclass[aps,pra,twocolumn,showpacs,groupedaddress]{revtex4-1}

\usepackage{hyperref}  
\hypersetup{
pdftitle={Dimer of two bosons in a one-dimensional optical lattice},
pdfauthor={Juha Javanainen, Otim Odong, and Jerome C. Sanders},
colorlinks=true,
urlcolor=blue,
linkcolor=blue,
bookmarks=true,
pdftoolbar=true,
filecolor=red,
citecolor=blue
}
\usepackage{graphicx}  
\usepackage{dcolumn}   
\usepackage{bm}        
\usepackage{amssymb}   
\usepackage{verbatim}  
\usepackage{epstopdf}
\usepackage{color}
\newcommand{\ket}[1]{\ensuremath{\left| #1 \right\rangle}}
\newcommand{\bra}[1]{\ensuremath{\left\langle #1 \right|}}
\newcommand{\braket}[2]{\ensuremath{\left\langle #1 \right|\left. #2
\right\rangle}}
\newcommand{\mele}[3]{\ensuremath{\left\langle #1 \right|#2\left| #3
\right\rangle}}
\newcommand{\dyad}[2]{{\ket{#1}\!\!\bra{#2}}}

\newcommand{\abs}[1]{\ensuremath{\left| #1 \right|}}

\newcommand{\beq}{\begin{equation}}
\newcommand{\eeq}{\end{equation}}
\newcommand{\bea}{\begin{eqnarray}}
\newcommand{\eea}{\end{eqnarray}}

\newcommand{\eq}[1]{{(\ref{#1})}}
\newcommand{\commentout}[1]{{}}

\newcommand{\half}{{\hbox{$\frac{1}{2}$}}}
\newcommand{\Kappa}{{\cal K}}

\definecolor{red}{rgb}{1,0,0}

\begin{document}
\title{Dimer of two bosons in a one-dimensional optical lattice}
\author{Juha Javanainen}
\affiliation{Department of Physics, University of Connecticut, Storrs, Connecticut 06269-3046}

\author{Otim Odong}
\affiliation{Department of Physics, University of Connecticut, Storrs, Connecticut 06269-3046}

\author{Jerome C. Sanders}
\affiliation{Department of Physics, University of Connecticut, Storrs, Connecticut 06269-3046}

\begin{abstract}
\commentout{
We investigate an exact solution for two bosons in a finite one-dimensional optical lattice using the Bose-Hubbard model.  The Hamiltonian includes tunneling from one site to the next and on-site interactions.  We have found analytic expressions for the eigenvalues and state vectors in momentum and position representations for a finite lattice.  These analytic solutions for bosons in a finite lattice represent the only solutions we know of that do not resort to limiting cases.  The general solutions are then investigated in several limiting cases and match those found by Valiente and Petrosyan \cite{ValienteM41} and Winkler et al.~\cite{WinklerK441}.  We also provide two examples for detection of our results, the size of the bound state and the transition rate.
}
We investigate theoretically the stationary states of two bosons in a one-dimensional optical lattice within the Bose-Hubbard model.  Starting from a finite lattice with periodic boundary conditions, we effect a partial separation of the center-of-mass and relative motions of the two-atom lattice dimer in the lattice momentum representation, and carefully analyze the eigenstates of the relative motion.  In the limit when the lattice becomes infinitely long, we find closed-form analytic expressions for both the bound state and the dissociated states of the lattice dimer. We outline the corresponding analysis in the position representation. The results are used to discuss three ways to detect the dimer:~by measuring the momentum distribution of the atoms, by finding the size of the molecule with measurements of atom number correlations at two lattice sites, and by dissociating a bound state of the lattice dimer with a modulation of the lattice depth.
\end{abstract}



\maketitle

\section{Introduction}

Optical lattices containing quantum degenerate Bose and Fermi gases~\cite{MOR06} have been a major topic in atomic, molecular, and optical physics of late, one motivation being experimental realizations of long-standing lattice models in condensed-matter physics and statistical mechanics~\cite{LEW06, BLO08}. In an optical lattice the transition amplitude (hence, probability) for an atom to tunnel from one lattice site to the next can be tuned over a wide range by adjusting the intensity of the lattice light. Feshbach resonances also permit a precise broad-range adjustment of the atom-atom interactions~\cite{FAT08, POL09}. This means that both the effective mass of the atoms and the atom-atom interactions are subject to experimental control.

Our broad theme is aggregates of atoms--call them lattice molecules--that can be formed and controlled in an optical lattice by controlling both the one-particle properties and the atom-atom interactions. The simplest one is the composite of two atoms, which we and others term a ``lattice dimer.'' Literature on variations of the lattice dimer is accumulating steadily (see~\cite{FED04, ORS05, WinklerK441, GRU07, NYG08FES, WAN08, NygaardN78, VAL08EPL, ValienteM41, PII08DIF, ValienteM42, ValienteMX} and references therein) and the trimer problem has also been addressed~\cite{VAL09THR}. It seems to us, however, that in the general quest toward novel physics seemingly elementary aspects such as the stationary states of the dimer and their limits when the lattice becomes infinitely long have been overlooked. The purpose of the present article is to remedy the situation.

We analyze the eigenstates of the Bose-Hubbard Hamiltonian for two bosonic particles in a lattice. We start in Sec.~\ref{FINITELATTICE} with a finite lattice, employing periodic boundary conditions. We transform the Hamiltonian into lattice momentum representation and separate what are the lattice analogs of the center-of-mass motion and relative motion of the atoms in a two-atom lattice dimer. The separation is, however, not complete~\cite{ORS05,GRU07,ValienteM41}: In the analysis of the relative motion, the amplitude of the tunneling between lattice sites will get scaled by a factor that depends on the center-of-mass motion. We, of course, have a numerical solution of the time-independent Schr\"odinger equation on hand, but also discuss qualitative features of the solution analytically. One bound state is found regardless of whether the atom-atom interactions are attractive or repulsive~\cite{WinklerK441}, along with what is the finite-lattice analog of the dissociation continuum of the lattice dimer. A surprising number of mathematical issues arise from the discreteness of the lattice and the boson symmetry of the states. We clarify them carefully.

The limit of a long lattice is the subject of Sec.~III\@. The main item here is the development of a closed-form analytic expression for the dissociated state of the dimer in the lattice momentum representation. Again, in contrast to past discussions in terms of Green's functions or scattering theory, we are after stationary states of the time-independent Schr\"odinger equation. Many past publications analyze the limit of a long lattice, occasionally without ever saying so. Inasmuch as a direct comparison is possible, our results agree with the previous results~\cite{WinklerK441,ValienteM41}.

Section~IV\@ presents the analysis of both the finite lattice and the long-lattice limit in the position (lattice site) representation~\cite{ValienteM41}. We reaffirm the results found from the momentum representation, and also find additional results such as an analytical approximation to the energies of the would-be dissociated states when there are many lattice sites.

As an example of the utility of our analytical results, we discuss  possible ways of detecting the lattice dimer in Sec.~V\@. Momentum distribution of the atoms~\cite{WinklerK441}, pair correlations of atom positions, and dissociation rate of a bound dimer when the depth of the lattice is modulated~\cite{WinklerK441} are the specific cases. The remarks in Sec.~VI conclude the article.

\section{Finite Lattice}\label{FINITELATTICE}

Let us start with the Bose-Hubbard model, ostensibly in one dimension, though the same mathematics apply with straightforward modifications in multiple dimensions. The Hamiltonian with standard conventions reads
\beq
\frac{H}{\hbar} \! = \! - \frac{J}{2} \sum_k \left(a^\dagger_{k+1} a_k + a^\dagger_{k-1} a_k \right) + \frac{U}{2} \sum_k a^\dagger_k a^\dagger_k a_k a_k \, .
\eeq
The index $k$ runs over the lattice sites, $L$ of them; $k=0,\ldots,L-1$. We use periodic boundary conditions, so that $k=L$ is the same as $k=0$, and likewise $k=-1$ means $k=L-1$. This  would be physically valid for a ring lattice. We do not intend to restrict the discussion to ring lattices, but the motivation is twofold. First, in a long-enough lattice the results will obviously be similar for arbitrary boundary conditions. Second, we have a sound framework to go mathematically to the limit of an infinitely long lattice.

The periodic boundary conditions bring in some subtle topology that we will not address but rather hide with our subsequent choices of the parameters. The principal one is that the number of lattice sites $L$ is taken to be even. This, again, is something that cannot materially influence the results for a very long lattice.  The sign of the site-to-site hopping amplitude $J$ is a matter of choice and can be flipped with the trivial canonical transformation $a_k\rightarrow (-1)^k a_k$. Here we take the native hopping amplitude to be positive, $J >0 $. A similar freedom does not apply to the strength of atom-atom interactions at each site; $U > 0$ and $U < 0$ correspond to repulsive and attractive interactions between the atoms, respectively.

In order to take advantage of the translation invariance, we next convert to momentum representation. The operators
\begin{equation}
c_q = \frac{1}{\sqrt{L}} \sum_k e^{-iqk} a_k
\label{PDEF}
\end{equation}
are also boson operators when $q$ runs over a suitable set of values, for instance, $q = 2\pi Q/L$ with the integers $Q$ picked so that we have $L$ quasimomenta in the first Brillouin zone of the lattice. For an even number of sites $L$ our standard choice is to allow the values $Q = -L/2+1,-L/2+2,\ldots,L/2$; $L$ of them. The addition of quasimomenta $q$ is understood modulo $2\pi$, so that the result belongs to the same set. The standard discrete Fourier transform relations
\begin{equation}
\sum_k e^{i q k} = L \delta_{[q,0]} \, , \quad \sum_q e^{i q k} = L \delta_{[k,0]}\,,
\end{equation}
where the brackets remind us of the fact that comparison of quasimomenta is modulo $2\pi$ and comparison of lattice sites modulo $L$, give the inverse of the definition~\eq{PDEF}
\begin{equation}
a_k = \frac{1}{\sqrt L}\sum_q e^{i q k}c_q\,.
\end{equation}
The Hamiltonian in momentum representation then becomes
\begin{eqnarray}
\frac{H}{\hbar}  =  \sum_q \omega_q c^\dagger_q c_q + \frac{{U}}{2L} \sum_{q_1,q_2,q_3,q_4} \delta_{[q_1+q_2,q_3+q_4]} c^\dagger_{q_1} c^\dagger_{q_2}c_{q_3}c_{q_4}, \nonumber \\
\label{FIRSTHAM}
\end{eqnarray}
where the transformation has diagonalized the term in the Hamiltonian describing site-to-site hopping,
\begin{equation}
\omega_q = - J \cos q \, . \label{omegaq}
\end{equation}

We study the most general state vector for two bosons,
\begin{equation}
| \psi \rangle = \sum_{p_1,p_2} A(p_1,p_2) \, c^\dagger_{p_1} c^\dagger_{p_2} | 0 \rangle \, .
\label{ANSATZOLD}
\end{equation}
Here $p_1$ and $p_2$ again are lattice momenta from the first Brillouin zone, and $\ket0$ is the particle vacuum. As the two creation operators commute, we assume without any loss of generality the symmetry of the $A$ coefficients $A(p_1,p_2)=A(p_2,p_1)$. Action of the atom-atom interaction part in the Hamiltonian then leads to
\bea
&& \sum_{q_1,q_2,q_3,q_4} \delta_{[q_1+q_2,q_3+q_4]} \,
c^\dagger_{q_1} c^\dagger_{q_2} c_{q_3} c_{q_4} \ket \psi \nonumber \\
& = & 2 \sum_{q_1,q_2,p_1,p_2} A(p_1,p_2) \, \delta_{[q_1+q_2,p_1+p_2]} \, c^\dagger_{q_1} c^\dagger_{q_2} \ket 0 \, ;
\label{ACTINTER}
\eea
see the Appendix. The atom-atom interaction therefore conserves the total lattice momentum of the atom pair, modulo $2 \pi$.

Suppose the state vector is confined to the subspace with the total momentum $P$ and write $p_{1,2} = \half P \pm q$. For clarification, we note that we refer to $p_1$, $p_2$, and $q$ as quasimomenta and $P$ as the total momentum even though all of the quantities are  obviously dimensionless.  This notation introduces a set of lattice-momentum-like quantities $q$ to describe the relative motion of the two atoms in such a way that the state~\eq{ANSATZOLD} reads
\beq
\ket \psi = \sum_q A(q) \, c^\dagger_{\half P + q} c^\dagger_{\half P -q} \ket 0 \, ,
\label{ANSATZ}
\eeq
where $A(q)$ now has the symmetry
\beq
A(q)=A(-q)\,.
\label{SYMMA}
\eeq
As always, the sums $\half P\pm q$ are modulo $2\pi$ so that they have the proper values of a lattice momentum in the first Brillouin zone.

There are two details to consider. First, as Eq.~\eq{ACTINTER} shows, the comparison of the total momenta $P$ in atom-atom interactions is modulo $2\pi$. Two particles at the edge of the Brillouin zone can scatter into two particles at the center of the Brillouin zone. On the other hand, $\half P$ stands for the average momentum of the two atoms, and two atoms at the center of the Brillouin zone ($P=0$) clearly have a different physical signature than two atoms at the edge of the Brillouin zone ($P=2\pi$). The proper range of $P$ is therefore $(-2\pi,2\pi]$. Second, the newly introduced summation index $q$ may have some peculiar properties. If $\half P$ in itself is a legal lattice momentum, then so is $q$, and the sum in~\eq{ANSATZ} is taken to run over the first Brillouin zone as usual. In contrast, if $\half P$ is not a lattice momentum,  the summation indices $q$ are of the form $q=2\pi Q/L$ with a {\em half-integer\/} $Q$, and the sum is taken to run over the values $Q = -L /2+1/2, \ldots,  L/2-1/2$. We then say that the relative motion is governed by half-integer lattice momenta.

At least in principle, a state with a half-integer lattice momentum may be prepared experimentally. Suppose that the atom-atom interactions are first turned off, for instance, by means of a Feshbach resonance, and two atoms are put in the states with $q_1=0$, $q_2=2\pi/L$. This is a state with a half-integer relative momentum. Moreover, if subsequently the atom-atom interactions are turned on so that the lattice translation invariance is not broken in the process, a nontrivial half-integer state is liable to arise for the relative motion of the two atoms. In the limit of an infinitely long lattice the difference between ordinary and half-integer lattice momenta must be irrelevant, but for completeness we occasionally mention it.

Let us continue with an ansatz of the form~\eq{ANSATZ}. In view of Eq.~\eq{ACTINTER}, the action of the Hamiltonian on the state vector will give
\begin{eqnarray}
\frac{H}{\hbar} | \psi \rangle & = & \sum_q \left[ \left( \omega_{\half P + q} + \omega_{\half P - q} \right) A(q) + \frac{U}{L} \sum_{q'} A(q') \right] \nonumber \\ && \times \, c^\dagger_{\half P + q} c^\dagger_{\half P - q} | 0 \rangle \, ,
\label{OPEXP}
\end{eqnarray}
where $q'$ is a legal or a half-integer lattice momentum exactly like $q$.
The time-independent Schr\"odinger equation $H|\psi\rangle = E|\psi\rangle$
is obviously satisfied if the coefficients of the products of the boson creation operators are the same for each $q$ on both sides of the equation, which is the case if the coefficients $A(q)$ satisfy
\begin{equation}
\left( \omega_{\half P + q} + \omega_{\half P - q} \right) A(q) + \frac{U}{L} \sum_{q'} A(q') = \frac{E}{\hbar} A(q) \, .
\label{TOSOLVE}
\end{equation}

Under the restriction~\eq{SYMMA}, Eq.~\eq{TOSOLVE} is not only sufficient but also necessary for the solution of the Schr\"odinger equation. This could be shown by restricting the summation index $q$ in the ansatz~\eq{ANSATZ} to non-negative values only. The ambiguity in the argument that the states $c^\dagger_{\half P +q}c^\dagger_{\half P -q}\ket0$ are the same for $q$ and $-q$ would thereby be removed, but at the cost of severe complications elsewhere. We therefore retain the double counting of the states in the formal analysis. So far we have restricted the values of the summation indices $q$ and $q'$ to the first Brillouin zone. However, if convenient, $A(q)$ may be regarded as a periodic function of $q$ with the period of $2\pi$. In that case the sums run over appropriate discrete values within a period, including an end point of the period only once.

Unlike for free atoms with a quadratic dispersion relation when the total momentum simply provides an additive constant to the energy, here the total momentum--center-of-mass momentum if you will--has a profound effect on the stationary states of the relative motion of the two atoms. Using \eq{omegaq} we have
\beq
\omega_{\half P+q}+\omega_{\half P-q} = -2 J \cos \left( \half P \right) \, \cos q \, , \label{OMEGAHALF}
\eeq
so that the center-of-mass motion couples to the relative motion described by the quasimomenta $q$ and effectively scales the hopping amplitude $J$. The center-of-mass motion and the relative motion do not completely separate~\cite{ORS05}.  For future use we define the frequency scale of the relative motion of the atoms for a given total center-of-mass quasimomentum $P$ as
\beq
\Omega_P \equiv 2 J \cos \left( \half P \right) \, .
\eeq
We also scale the dimensional parameters in the problem to $\Omega_P$, defining the dimensionless quantities representing energy of a state and strength of the atom-atom interactions as follows:
\beq
\omega \equiv \frac{E}{\hbar \Omega_P} \, , \quad \Kappa \equiv \frac{{U}}{\Omega_P} \, .
\label{DIMLESS}
\eeq

With these ingredients the solution of the time-independent Schr\"odinger equation is reduced to two steps. First, Eq.~\eq{TOSOLVE} gives
\beq
A(q) = \frac{\frac{{U}}{L} \sum_{q'} A(q')}{E/\hbar+ 2 J \cos(\half P)\cos q}\,,
\label{WFEQ}
\eeq
and this further gives the equation for the energy eigenvalue
\begin{equation}
\frac{{U}}{L}\sum_q\frac{1}{E/\hbar+ 2 J \cos(\half P)\cos q} = 1\,.
\label{SCH1}
\end{equation}
In the dimensionless variables~\eq{DIMLESS}, Eq.~\eq{SCH1} reads
\beq
f(\omega,L)=\frac{1}{L} \sum_q \frac{1}{\omega + \cos q} = \frac{1}{\Kappa} \, .
\label{ENEQ}
\eeq

Whether we deal with integer or half-integer quasimomenta, we may always pair them up in the sum in Eq.~\eq{SCH1} so that for every $q$ there is a $q'$ such that $\cos q = -\cos q'$. This implies that the sign of the factor $2 J \cos(\half P)$ has no effect on the possible solutions $E$ of Eq.~\eq{SCH1}, that is, on the energy spectrum. The energy eigenvalues are the values of $E$ that satisfy \eq{SCH1}.  The problem can also be discussed in terms of the dimensionless scaled energy $\omega$.  In this view, the solutions are the values of $\omega$ that satisfy \eq{ENEQ}.  Using either of these two equations, it is easy to solve the energy spectrum numerically for a wide range of parameters. From this point onward, unless otherwise stated or implied, we employ the dimensionless scaled energy $\omega$.

\begin{figure}[b]
\includegraphics[width=8.0cm]{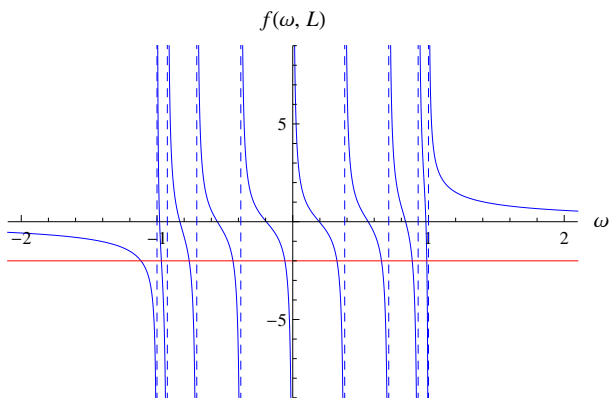}
\caption{(Color online) The function $f(\omega,L)$ [Eq.~\eq{ENEQ}] for $L=16$. The horizontal line represents the right-hand side of Eq.~\eq{ENEQ} for $\Kappa = - \half$. The dashed vertical lines are the asymptotes of $f(\omega,L)$ at the values of $\omega=-\cos q$ such that $f(\omega,L)=\pm \infty$.}
\label{FFIG}
\end{figure}

A standard discussion ensues from Eq.~\eq{ENEQ} when we plot $f(\omega,L)$ as a function of $\omega$. In Fig.~\ref{FFIG} the lattice has $16$ sites, $L = 16$, and the added horizontal line depicts the value of the right-hand side of Eq.~\eq{ENEQ} for $\Kappa = - \half$. It is clear that $f(\omega,L)$ may attain any finite value between two successive arguments $\omega = - \cos q$ for the given discrete set of quasimomenta $q$.  Vertical asymptotes (dashed lines) are plotted in Fig.~\ref{FFIG} at each value of $\omega = -\cos q$. For $L = 16$ there are eight regions bounded by such asymptotes, obviously because $\cos(q) = \cos (-q)$. This does not signal doubly degenerate energies $\omega$, but is one manifestation of the fact that we have counted the basis vectors with $q \ne 0$ and $q \ne \pi$ twice.

For a noninteracting system with ${\Kappa}\rightarrow 0$, the left-hand side of Eq.~\eq{ENEQ} approaches plus or minus infinity depending on the sign of the interaction parameter ${U}$, and the spectrum of the scaled energies is then precisely the numbers $\omega = - \cos q$.  This is the finite-lattice analog of the continuum of the relative motion of two noninteracting atoms.  The essential twist is that the width of the continuum band of energies $\Delta E = 2\hbar\Omega_P$ depends on the center-of-mass momentum of the atom pair $P$.

Let us next restore the atom-atom interactions so that $1/|\Kappa|<\infty$. The scaled energies $\omega$ that resided between the values of $-\cos q$, in the band with $-1 < \omega < 1$, will stay this way.  These states are the finite-lattice analog of the dissociation continuum of the lattice dimer.  For brevity, we call the states with $-1 < \omega < 1$ continuum states, even though there are never true continuum states in any finite-length lattice.  The dimensionless energy values for the continuum states are labeled $\omega_c$.

Moreover, as may be seen from Fig.~\ref{FFIG}, there is one energy value that moves away from the dissociation continuum when $|{\Kappa}|$ increases and therefore $1/|\Kappa|$ decreases.  This is obviously the bound state of the lattice molecule.  The bound state is characterized by $|\omega| > 1$, and we label the corresponding dimensionless energy value $\omega_b$.  The unscaled energy of the bound state $E = \hbar \Omega_P  \omega_b$  also varies with the center-of-mass motion, another feature that does not exist in free space. For attractive interactions, $U<0$, the bound state is the lowest-energy state of the lattice dimer and lies below the dissociation continuum. However, for repulsive interactions, $U>0$, the bound molecular state lies above the dissociation continuum, a situation that also has no analog for molecules in free space. The bound state above the continuum is a manifestation of the trivial symmetry of the Hamiltonian, $H(J,{U}) = - H(-J,-{U})$, combined with the observation that we have already made that the transformation  $J \rightarrow -J$ alone has no effect on the spectrum of energy eigenstates.

Having discussed the energy spectrum, we next determine the state vectors.  The generic form is specified by Eq.~\eq{ANSATZ}.  The coefficients $A(q)$ are given by Eq.~\eq{WFEQ}, where the numerator is just a constant and will eventually be absorbed into the normalization. There is another subtlety here. Namely, in quantum mechanics one wants to do explicit calculations using expansion coefficients with respect to an orthonormal basis, but  in the present case the states $c^\dagger_{\half P +q} c^\dagger_{\half P -q}\ket{0}$ are not orthonormal for all $q$.  The states corresponding to $q$ and $-q$ are the same, and moreover the states with $q=0$ and $q=\pi$, which occur for integer quasimomenta, are normalized to two, not one.  All of this requires a careful documentation of the range of the summation index $q$ and, depending on the range used in the calculations, may lead to peculiar factors in the expressions governing the normalization of the state.  The true dimension of the state space is $L/2+1$ if we deal with integer quasimomenta and $L/2$ for half-integer quasimomenta.

In our subsequent calculations we continue the double counting but normalize the expansion coefficients so that the underlying quantum states are always normalized in the standard way. Now, the inner product of two states of the form~\eq{ANSATZ} with the expansion coefficients $A(q)$ and $B(q)$ as inherited from quantum mechanics is
\begin{eqnarray}
\braket{\psi_A}{\psi_B}\! & = & \! 4 \!\!  \sum_{0<q<\pi} \!\!\! A^*(q) B(q)\! +\! 2 A^*(0)B(0) \!+ \!2A^*(\pi)B(\pi)  \nonumber \\ & = &
\! 2 \!\!\!\! \sum_{-\pi < q \le \pi} \!\!\!\!\! A^*(q)B(q) \, ;
\end{eqnarray}
if $q$ is a half-integer quasimomentum, then the last two terms in the second form are missing, but the final result again holds true. We therefore define the inner product for the expansion coefficients,
\beq
(A,B) = 2  \!\!\!\! \sum_{-\pi < q \le \pi} \!\!\!\!\! A^*(q)B(q) \, ,
\label{MODSCP}
\eeq
and normalize accordingly. Orthonormality with respect to the inner product~\eq{MODSCP} is equivalent to orthonormality of the underlying quantum states.  Since the version of the Schr\"odinger equation~\eq{TOSOLVE} is Hermitian with respect to this inner product, the true quantum states come out with the proper orthonormality properties as well.

In view of the eigenvalue equation~\eq{WFEQ} and the definition of the dimensionless variables \eq{DIMLESS}, the eigenstate for the eigenvalue $\omega$ is defined by the expansion coefficients
\beq
A(\omega,q) = \frac{C(\omega)}{\omega+\cos q} \,.
\label{EIGSTA}
\eeq
Moreover, the choice of the overall numerical factor
\beq
C(\omega) = \left[ \sum_{q} \frac{2}{(\omega+\cos q)^2} \right]^{-1/2}
\label{NORCON}
\eeq
ensures unit normalization with respect to the inner product~\eq{MODSCP}.

What these results mean in terms of site occupation numbers is seen by converting the momentum representation state~\eq{ANSATZ} back to lattice site (position) representation using Eq.~\eq{PDEF}. The result is
\beq
\ket{\psi(\omega)} = \frac{1}{L} \!\! \sum_{q,k_1,k_2} \!\!\!\! A(\omega,q) \, e^{\half i(k_1+k_2)P}e^{i(k_1-k_2)q}  a^\dagger_{k_1} a^\dagger_{k_2} \ket 0 \, .
\label{TRUEEIGNSTATE}
\eeq
For $P=0$ this state remains unchanged in lattice translations; for $P \ne 0$ the phase of the state changes by $e^{iP}$ when one moves the reference point $k=0$ back by one step in the lattice, meaning that $k_1\rightarrow k_1+1$ and $k_2\rightarrow k_2+1$.  The quantum mechanical stationary state of a molecule is evenly spread along the entire lattice. Lest this appear odd, the same applies to any aggregate of atoms also in free space. How the spread-out stationary states relate to our intuition that a molecule is a localized object will be demonstrated in Sec.~\ref{PAIRCORRELATIONS}.

At this point it is tempting to introduce the analogs of the center-of-mass position and the relative position for the two atoms, expressed in terms of the atomic positions $k_1$ and $k_2$ as
\beq
K = \half(k_1+k_2),\quad k = k_1-k_2\,.
\eeq
Formally, we write from Eq.~\eq{TRUEEIGNSTATE}
\begin{eqnarray}
\ket{\psi(\omega)} & = & \frac{1}{\sqrt{L}} \sum_{K} \Bigg{(} e^{iPK} \Bigg{\{} \sum_k \left[ \frac{1}{\sqrt{L}} \sum_q A(\omega,q) e^{iq k} \right] \nonumber \\ & & \times \; a^\dagger_{K+\half k} a^\dagger_{K - \half k}
\ket{0} \Bigg{\}} \Bigg{)} \, .
\label{EINSI}
\end{eqnarray}
However, this expression is incorrect, formal only, since the summation indices $K$ and $k$ do not decouple in a simple way. For instance, $K\pm \half k$ should be a legitimate lattice index, an integer. Equation~\eq{EINSI} appears to represent a superposition of copies of the molecule with the center of mass fixed at $K=0$,
\beq
\ket{\omega} = \sum_k
\left[\frac{1}{{\sqrt{L}}}\sum_q A(\omega,q) e^{iq k} \right]
a^\dagger_{\half k}a^\dagger_{-\half k} \ket{0} \, ,
\label{EINSIND}
\eeq
translated along the lattice to each site $K$, although problems with the summation indices persist.

This unsuccessful attempt is as close as we have got to separating the center-of-mass and relative motions in second-quantized position representation. Nonetheless, we qualitatively think of the expression inside the square brackets in Eq.~\eq{EINSIND} as the wave function of the relative motion of the atom pair.

\section{Limiting Case of a Long Lattice}

While it is easy to solve both the energy spectrum and the state vectors numerically for a wide range of parameters, some analytical work is also possible in the limit of a large number of lattice sites. In the limit $L\rightarrow\infty$ the quasimomenta $q$ make an infinitely dense set over the interval $(-\pi,\pi]$. Our main technical tool is the continuum approximation for the quasimomenta
\beq
\sum_q f(q) \simeq \frac{L}{2 \pi} \int_{-\pi}^\pi dq f(q) \, ,
\label{CONTLIM}
\eeq
valid for ``sufficiently smooth'' functions of quasimomentum  $f(q)$.

Now consider the would-be continuum states in the limit of an infinite number of lattice sites.  A simple argument can be found by examining Fig.~\ref{FFIG}.  As $L \rightarrow \infty$, the curves in the domain $-1 < \omega < 1$ get closer and closer together in such a way that they in effect become straight vertical lines.  Thus, the solutions for the continuum energies in the limit of a large number of lattice sites are given by the locations of the asymptotes that occur at
\beq
\omega_c(Q) = -\cos\left(\frac{2 \pi Q}{L} \right)\,.
\eeq
As before, we use the subscript $c$ as in $\omega_c$ to mark a consideration that is specific to the continuum states with $-1<\omega_c<1$, and likewise $b$ for the bound state with $|\omega_b|>1$.

The continuum energies become infinitely dense with $L\rightarrow\infty$. An attempt to isolate any individual energy and eigenstate eventually becomes futile, and all observable quantities must be expressible as sums over the energies $\omega_c$. Originally, we had the integer $Q$ run over the interval $(-L/2,L/2]$, but, as already noted, this range duplicates all but at most one energy eigenvalue. Given a ``sufficiently smooth'' function of the energy, $g(\omega)$, we therefore approximate
\bea
\sum_{\omega_c} g(\omega_c) &\simeq& \sum_{Q=1}^{L/2} g\left[-\cos\left(\frac{2 \pi Q}{L} \right) \right] \nonumber \\ & \simeq & \int_{1}^{L/2} dQ \, g\left[-\cos\left(\frac{2 \pi Q}{L} \right) \right] \nonumber \\ & \simeq & \frac{L}{2\pi} \int_{-1}^{1} \frac{d\omega_c}{\sqrt{1-\omega_c^2}} \, g(\omega_c) \, .
\label{ENCONLIM}
\eea
This expression identifies
\beq
\varrho(\omega_c) = \frac{L}{2\pi} \frac{1}{\sqrt{1-\omega_c^2}}
\label{DENSITYOFSTATES}
\eeq
as the density of continuum states, normalized to the dimensionless energy of the variable $\omega_c$.

Regardless of the ultimate limit of an infinitely long lattice, $L \rightarrow \infty$, we always think of the lattice as finite.  All inner products, normalizations, and so on, are with respect to discrete sums.  When appropriate, these sums are just approximated as in Eqs.~\eq{CONTLIM} and~\eq{ENCONLIM}.

\subsection{Bound state}

First we will analyze the energy and the normalization of the bound state.  In the limit of a large number of lattice sites, we approximate
\bea
\frac{1}{L} \sum_q\frac{1}{\omega_b +\cos q}
 & \simeq &\frac{L}{2\pi} \frac{1}{L}\int_{-\pi}^{\pi} dq \, \frac{1}{\omega_b + \cos q} \nonumber \\ & = & \frac{{\rm sgn} \, (\omega_b)}{\sqrt{\omega_b^2 - 1}} \, , \label{CAPPROXBS}
\eea
a valid process since for any $\Kappa\neq0$ the bound state has $| \omega_b| > 1$ and the function of $q$ to be summed does not become singular in the limit $L\rightarrow\infty$.  We therefore find the bound-state energy by substituting \eq{CAPPROXBS} in \eq{ENEQ} and solving for $\omega_b$,
\beq
\omega_b = {\rm sgn}(\Kappa) \sqrt{1+ \Kappa^2}\,. \label{BSOMEGA}
\eeq
In the limit of strong atom-atom interactions, this reduces to $\omega_b \simeq \Kappa$ and so in dimensional units the energy of the bound state is $E_b = \hbar {U}$, but while $|\Kappa| \sim 1$ the energy of the bound state depends on the center-of-mass motion.  These results for the bound-state energy agree with those found, for example, in~\cite{WinklerK441} and~\cite{ValienteM41}, although these authors do not mention the limit $L\rightarrow\infty$.

The state vector for the bound state is of the form of \eq{EIGSTA} and \eq{NORCON} with $\omega = \omega_b = \pm \sqrt{1 + \Kappa^2}$. To find the normalization coefficient in the limit $L \rightarrow \infty$, we calculate
\beq
[C_{b}(\omega_b)]^{-2}  \simeq  \frac{2L \abs{\omega_b} }{(\omega_b^2 - 1)^{3/2}} \, . \label{COMEGAB}
\eeq
The bound state is specified by \eq{EIGSTA}, \eq{COMEGAB}, and \eq{BSOMEGA} as
\beq
A_{b}(\omega_b,q) = \frac{|{\Kappa}|^{3/2}}{\sqrt{2 L |\omega_b|}}\,\frac{1}{\omega_b + \cos q}\,.
\label{BSTATE}
\eeq

\subsection{Continuum states}

We next consider the state vectors for the continuum states.  While the bound state has a straightforward limit as $L \rightarrow \infty$, the continuum states with $-1 < \omega_c < 1$ pose a problem.  For any finite number of lattice sites the stationary state is, of course, still of the form
\beq
A_c(\omega_c,q) = \frac{C_c(\omega_c)}{\omega_c + \cos q}\,.
\label{ONCEMORE}
\eeq
However, as the continuum states get denser with $L \rightarrow \infty$, this expression becomes singular as a function of $q$ in the neighborhood of $\omega_c+\cos q\simeq 0$, and a rule for handling the singularity has to be defined.

Our assignment is basically the following: In principle we only do discrete sums, and knowing the exact discrete continuum eigenvalues $\omega_{c}$ would always allow us to carry out the sums without problems.  Nevertheless, in practice we would like to approximate the sums as integrals, as in~\eq{CONTLIM}. We are therefore looking for a function of $q$, $A_{c}(\omega_{c}, q)$, which would give the same value from a continuum-limit integral~\eq{CONTLIM} as does a discrete sum involving the amplitudes~\eq{ONCEMORE}. We will, in fact, find such a function $A_{c}(\omega_{c}, q)$  [Eq.~\eq{CSTATE}]. Moreover, it comes with the appealing property that, unlike in applications of Eq.~\eq{ONCEMORE}, one does not have to know the energy $\omega_c$ precisely.

Following Ref.~\cite{FanoU124}, we try the following ansatz:
\beq
\frac{1}{\omega_{c}+\cos q} = \textrm{P}\frac{1}{\omega_{c}+\cos q} + \Pi(\omega_{c})\delta(\omega_{c}+\cos q)\,,
\label{BREAKUP}
\eeq
where $\textrm{P}$ stands for principal-value integral and $\Pi(\omega_{c})$ is a function to be determined. To find $\Pi(\omega_{c})$, we attempt to solve the continuum version of the Schr\"odinger equation itself. By rearranging Eq.~\eq{WFEQ} and using the dimensionless variables \eq{DIMLESS}, it is
\beq
(\omega_{c} + \cos q) A_c(q,\omega_{c})  =  \frac{\Kappa}{2\pi}\int_{-\pi}^\pi dq' \, A_c(\omega_{c},q') \, .
\label{NEWSCHREQ}
\eeq
Substituting Eq.~\eq{ONCEMORE} into~\eq{NEWSCHREQ} and taking into account that for $\omega_{c}\in(-1,1)$ we have
\bea
&& (\omega_{c}+\cos q) \, \textrm{P} \frac{1}{\omega_{c} + \cos q} = 1 , \\
&& (\omega_{c}+\cos q) \, \delta(\omega_{c} + \cos q) = 0 , \\
&& \textrm{P} \int_{-\pi}^\pi dq \, \frac{1}{\omega_{c} + \cos q} = 0 \, ,
\label{PVINT}
\eea
we immediately find that
\beq
\Pi(\omega_{c}) = \frac{\pi\sqrt{1-\omega_{c}^2}}{\Kappa}
\label{GFORM}
\eeq
leads to the eigenvalue $\omega_c$ in Eq.~\eq{NEWSCHREQ}.

Next we digress on the normalization of the continuum wave functions. In the original discrete case two state vectors corresponding to two different energies are orthonormal in the sense of a Kronecker $\delta$,
\beq
2\sum_q A_c(\omega_{c},q)A_c(\omega_{c}',q) = \delta_{\omega_{c},\omega_{c}'}\,.
\eeq
Taking an ``arbitrary'' function $F(\omega_{c})$, we therefore have
\beq
\sum_{\omega_{c}'}\sum_q A_c(\omega_{c},q) A_c(\omega_{c}',q) F(\omega_{c}') =  \half\, F(\omega_{c})\,.
\label{DOUBLESUM}
\eeq
As one more remnant of the state counting problems, there are only half as many energies $\omega_c$ as there are coefficients $A_c(\omega_c,q)$. The sum over $\omega_c'$ is to be understood accordingly.
On the other hand, we have
\bea
\sum_{\omega_{c}} F(\omega_{c})  \simeq   \frac{L}{2\pi}\int_{-1}^{1} \frac{d\omega_{c}}{\sqrt{1 - \omega_{c}^2}} \, F(\omega_{c}) \, .
\eea
Using the continuum approximation for the sum over $q$ in Eq.~\eq{DOUBLESUM} once more, we have
\bea
\left(\frac{L}{2\pi}\right)^2\int_{-1}^{1} \frac{d\omega_{c}'\,F(\omega_{c}')}{\sqrt{1-\omega_{c}'^2}}
\int_{-\pi}^{\pi}dq \, A_c(\omega_{c},q) A_c(\omega_{c}',q) & & \nonumber \\ = \half F(\omega_{c}) = \half \int_{-1}^1 d\omega_{c}' \, \delta(\omega_{c}-\omega_{c}') F(\omega_{c}') \, . & &
\eea
The correct continuum approximation normalization of the coefficients $A_c(q,\omega_{c})$ therefore reads
\nopagebreak[1]
\bea
& & \int_{-\pi}^{\pi} dq \, A_c(\omega_{c},q) A_c(\omega_{c}',q)  \nonumber \\ & & = \half \sqrt{1 - \omega_{c}'^2} \left( \frac{2\pi}{L} \right)^2 \delta(\omega_{c} - \omega_{c}') \, .
\label{NORMREQ}
\eea

To implement this normalization, we start from Eqs.~\eq{ONCEMORE} and~\eq{BREAKUP}, substitute $x=-\cos q$, and find
\begin{widetext}
\beq
\int_{-\pi}^{\pi} \, dq A_c(\omega_{c},q) A_c(\omega_{c}',q)
= 2 C_c(\omega_{c}) C_c(\omega_{c}') \! \int_{-1}^{1} \! \frac{dx}{\sqrt{1-x^2}} \! \left[\textrm{P} \frac{1}{\omega_{c}\! -\! x} \! + \! \Pi(\omega_{c})\delta(\omega_{c}\! -\! x)\right] \!\! \left[\textrm{P}\frac{1}{\omega_{c}'\! -\! x}\!  +\!  \Pi(\omega_{c}')\delta(\omega_{c}'\! -\! x)\right] \, .
\eeq
By virtue of the form of the function $\Pi(\omega_{c})$, Eq.~\eq{GFORM}, the integrals of the products involving a principal value and a $\delta$ function cancel, and the integral of the product of two $\delta$ functions is simple. The product of two principal-value integrals can be handled with the identity~\cite{FanoU124}
\beq
\textrm{P}\frac{1}{\omega_{c}-x} \textrm{P}\frac{1}{\omega_{c}'-x}
= \frac{1}{\omega_{c}-\omega_{c}'}\left( \textrm{P}\frac{1}{\omega_{c}'-x} -\textrm{P}\frac{1}{\omega_{c}-x}\right) + \pi^2 \delta(x - \omega_{c}) \delta(x - \omega_{c}') \, .
\eeq
The first two terms both give zero by virtue of Eq.~\eq{PVINT}, so that we are left with
\beq
 \int_{-\pi}^{\pi} dq \, A_c(\omega_{c},q) A_c(\omega_{c}',q)
 = 2 C_c(\omega_{c})^2\, \frac{\pi^2 + \Pi(\omega_{c})^2} {\sqrt{1-\omega_{c}^2}} \, \delta(\omega_{c}-\omega_{c}') \, .
\eeq
Comparison with Eq.~\eq{NORMREQ} gives the normalization coefficient $C_c(\omega_{c})$, and hence the complete state vector for the continuum state
\bea
&&A_c(\omega_{c},q)  = \frac{\Kappa \sqrt{1-\omega_{c} ^2}}{L \sqrt{\Kappa^2+1-\omega_{c} ^2}} \Bigg{[} \textrm{P}\frac{1}{\omega_{c}+\cos q} + \frac{\pi\sqrt{1-\omega_{c}^2}}{\Kappa}
\delta(\omega_{c}+\cos q) \Bigg{]} \, .
\label{CSTATE}
\eea
\end{widetext}

The precise meaning of Eq.~\eq{CSTATE} should be borne in mind very carefully.  This is an approximation to the continuum wave function to be used when one replaces the sum over discrete quasimomenta with the continuum-limit integral as in Eq.~\eq{CONTLIM} and is purportedly valid in the limit of a large number of lattice sites, $L\rightarrow \infty$. At this point one, in principle, needs to know the continuum eigenvalue $\omega_{c}$. However, if the final result of the calculation varies smoothly with $\omega_{c}$, one may equally well think of $\omega_{c}$ as a continuous energy with the density of states~\eq{DENSITYOFSTATES}.

\commentout{
However, the normalization in this result is still unusual; the state~\eq{CSTATE} is basically the large-$L$ limit of a discrete eigenstate. For true continuum states one would rather normalize, say, to energy. Since we have up to now written all $q$ integrals with respect to the measure $dq L/2\pi$ that ultimately means that we are approximating discrete sums over discrete values of the $Q$ modes by integrals over $q$, we continue in this way. Then the condition for normalization to energy reads
\beq
\frac{L}{2\pi} \int dq\, A_c(\omega,q)A_c(\omega',q) = \delta(\omega-\omega').
\eeq
The corresponding normalized state vector is
\beq
A(\omega,q) = \frac{2 \pi \Kappa}{\sqrt{\Kappa^2+1-\omega ^2}}\left[ \textrm{P}\frac{1}{\omega+\cos q} + \frac{\pi\sqrt{1-\omega^2}}{\Kappa}
\delta(\omega+\cos q)\right]\,.
\label{CSTATE}
\eeq
}

\commentout{
We next ask what kind of an experiment could detect the peculiar properties of the lattice molecule. We begin with the energy spectrum, studied by means of electromagnetically induced transitions, say, rf transitions or two-photon optical transitions. To the leading order in the tight-binding approximation the atomic transitions only take place within each potential well separately, so that the transition operator between the initial and final states, annihilated by the operators $a_k$ and $b_k$, is schematically of the form
\beq
H'_0 = \xi_0 \sum_k (a^\dagger_k b_k+b^\dagger_k a_k)\,.
\eeq
Unfortunately, such a perturbation does not couple to the motion of the atoms between the wells, so that it will not dissociate the lattice molecule and cannot be employed to determine the spectrum.

In the next approximation there is some overlap between the wave function of the atoms in adjacent potential wells, so that there is a transition operator of the form
\beq
H'_1 = \xi_1 \sum_k (a^\dagger_k b_{k+1}+b^\dagger_{k+1} a_k)\,.
\eeq
This is of the same form as the hopping term, the analog of kinetic energy in the tight-binding approximation. If there are atom-atom interactions, this Hamiltonian is not diagonal in the motional states of the atoms and could be enlisted to study the spectrum of the molecule. Here, however, we consider as another example another scheme in which the amplitude of the driving electromagnetic field varies linearly along the lattice, so that the relevant transition operator is of the form
\beq
H'_2 = \xi_2 \sum_k k(a^\dagger_k b_{k}+b^\dagger_{k} a_k)\,.
\eeq

In the momentum representation this Hamiltonian will read
\bea
H'_2 &=& \xi_2 \sum_k k(b^\dagger_k a_{k}+{\rm h.c.})\nonumber\\
 &=& \frac{ \xi_2}{L} \sum_{p_1,p_2,k} k
\eea
AND SO ON... .
}

\section{Dimer in Position Representation}
\label{SOLPOSREP}

For comparison and additional results, we next study the lattice dimer in position representation, somewhat parallel to Ref.~\cite{ValienteM41}. The starting point is the time-independent Schr\"odinger equation~\eq{TOSOLVE} in the case when both $\half P$ and $q$ are legal quasimomenta. Some changes in the formulation would result for half-integer quasimomenta, but we do not discuss them.

As already noted in connection with the somewhat ill-defined equations~\eq{EINSI} and~\eq{EINSIND}, we use
\beq
\alpha_k = \frac{1}{\sqrt{L}} \sum_q e^{i q k} A(q)
\label{DISCFT}
\eeq
to represent the relative motion of the two atoms in position representation, $k$ being the distance in lattice units between the atoms. From another viewpoint, the step from $A(q)$ to $\alpha_k$ in Eq.~\eq{DISCFT} is a discrete Fourier transformation, a perfectly well-defined mathematical operation. The transformation preserves the inner product~\eq{MODSCP}, so that the position-representation states are normalized exactly as in momentum representation.

Transforming Eq.~\eq{TOSOLVE}, we directly have an equation for the amplitudes $\alpha_k$,
\beq
\omega\alpha_k + \half\alpha_{k+1} + \half\alpha_{k-1} =  \delta_{k,0} \,
\Kappa \alpha_0 \, .
\label{NEWEQ}
\eeq
Here we choose the values of the relative coordinate of the two atoms from the interval $k\in(-L/2,L/2]$, but more generally, $\alpha_k$ must be regarded as a periodic function of $k$ with the period $L$.

Now, Eq.~\eq{NEWEQ} is a second-order finite-difference equation that may be solved like the corresponding second-order differential equation. The homogeneous equation
\beq
\omega\alpha_k+\half\alpha_{k+1} +\half\alpha_{k-1} = 0
\label{HOMEQ}
\eeq
admits solutions of the form $\alpha_k = x^k$, where we see from direct substitution that the constant $x$ may take on the values
\beq
x = -\omega \pm \sqrt{\omega^2-1}\,.
\eeq
The product of the two possible values of $x$ is always unity.

\subsection{Bound state}
First consider the case $|\omega|>1$. In this case, $\omega =\omega_b$ and $x$ is a real number. Suppose $\omega_{b}>1$, then the two values of $x$ smaller and larger than unity in absolute value are
\beq
x_< = -\omega_{b}+\sqrt{\omega_{b}^2-1},\quad x_> = -\omega_{b}-\sqrt{\omega_{b}^2-1}\,.
\eeq
By the boson symmetry, we may always require that $\alpha_k = \alpha_{-k}$. Recalling that $x_> = x_<^{-1}$, the only possible solution to Eq.~\eq{NEWEQ} is then of the form
\beq
\alpha_k =
\left\{
\begin{array}{ll}
A x_<^k + B x_>^k,&k\ge 0\,,\\
A x_>^k + B x_<^k,&k\le 0\,.
\label{POSBND}
\end{array}
\right.
\eeq
Substitution into Eq.~\eq{NEWEQ} with $k=0$ gives
\beq
(A-B)\sqrt{\omega_{b}^2-1} = (A+B)\Kappa\,.
\label{EVALEQ1}
\eeq

However, there is another equation to reckon with. Namely, to satisfy the periodic boundary conditions, the solution must be periodic, so that at $k=L/2$ the expression \eq{POSBND} must switch between the $k\ge0$ and $k\le0$ forms while remaining a solution to \eq{HOMEQ}. This leads to a second condition,
\bea
&& \omega_{b} \left( A x_<^{L/2} + B x_>^{L/2} \right) + \half \left( A x_>^{-L/2+1} + B x_<^{-L/2+1} \right. \nonumber \\ && \,\,\,\,\,\, \left. + A x_<^{L/2-1} + B x_>^{L/2-1} \right) = 0 \, ,
\eea
or
\beq
B = A x_<^L\,.
\label{EVALEQ2}
\eeq
In the limit of an infinitely long lattice, $L\rightarrow\infty$, we have $B/A=0$, so that Eq.~\eq{EVALEQ1} gives
\beq
\omega_{b} = {\rm sgn}(\Kappa)\sqrt{1+\Kappa^2}\,.
\label{BENE}
\eeq
Here we have worked out the case $\Kappa<0$ as well, which leads to $\omega_{b} < -1$. The result, of course, is as before [Eq.~\eq{BSOMEGA}]. The state vector for the bound state is~\cite{ValienteM41}
\beq
\alpha_k(\omega_b) = A x_<^{|k|} = \sqrt\frac{\Kappa}{2 \omega_b} (\Kappa-\omega_b)^{|k|},
\label{POSWF}
\eeq
where we have explicitly normalized, $2\sum_k |\alpha_k|^2 = 1$.

If the lattice is not infinitely long, the bound state also has a component that grows exponentially away from the center site $k=0$ as $|x_>|^{|k|}$. Moreover, the bound-state energy is shifted from the infinite-lattice value. In fact, Eqs.~\eq{EVALEQ1} and \eq{EVALEQ2} have a solution for $A$ and $B$ if and only if
\bea
\sqrt{\omega_{b}^2-1}-\Kappa & = & x_<^L (\sqrt{\omega_{b}^2-1}+\Kappa) \nonumber \\ & = & \frac{\sqrt{\omega_{b}^2-1} + \Kappa}{(\sqrt{\omega_{b}^2-1} + \omega_{b})^L}\,.
\eea
If this equation for $\omega_{b}$ has an analytic solution for a general (even) $L$, we have not been able to find it. However, the equation poses no particular problems numerically, and we have verified that the solution agrees with what is found from Eq.~\eq{ENEQ}.
The coefficients $A$ and $B$ can be determined explicitly in terms of $\omega_{b}$ so that the state~\eq{POSBND} is properly normalized, but the best we have been able to do analytically is cumbersome. Generally speaking, if a finite lattice is the issue, we believe that there is little to be gained from an attempt to press analytical as opposed to numerical calculations.

\subsection{Continuum states}

Second, consider the case $|\omega|<1$, so that $\omega=\omega_c$. Then $x_<$ and $x_>$ are complex numbers of unit modulus, of the form $e^{\pm iq}$  for some real $q$;
\beq
\begin{array}{lccl}
x_< & = & -\omega_{c} + \sqrt{\omega_{c}^2-1} & = e^{iq} \, , \nonumber \\  x_> & = & -\omega_{c} - \sqrt{\omega_{c}^2-1} & = e^{-iq} \, .
\end{array}
\label{xgl}
\eeq
The solutions to Eq.~\eq{NEWEQ} with the requisite boson symmetry are
\beq
\alpha_k(\omega_{c}) = \left\{
\begin{array}{ll}
A \cos kq + B \sin kq,& k\ge0;\\
A\cos kq - B \sin kq,& k\le0\,.
\label{ANSCON}
\end{array}
\right.
\eeq
From Eqs.~\eq{xgl}, the corresponding energy is of the form
\beq
\omega_{c} = -\cos q\,.
\label{CENVAL}
\eeq
Equation \eq{NEWEQ} with $k=0$ then gives
\beq
B \sin q = \Kappa A\,.
\eeq
Once more, the ansatz \eq{ANSCON} must give a solution to Eq.~\eq{HOMEQ} also at $k=L/2$, where the solution switches between the two forms, so that we have
\bea
&& \omega_{c} \left\{A \cos \left[ \left( \half L \right) q \right] \! + \! B \sin \left[ \left( \half L \right) q \right] \right\} \nonumber \\ && \! + \! \left\{ A \cos \left[ \left( \half L - 1 \right) q \right] \! + \! B \sin \left[ \left( \half L - 1 \right) q \right] \right\} = 0 \, ,
\eea
or
\beq
B = A \tan\frac{Lq}{2}\,.
\label{CCOND}
\eeq

\begin{figure}[t]
\includegraphics[width=8.0cm]{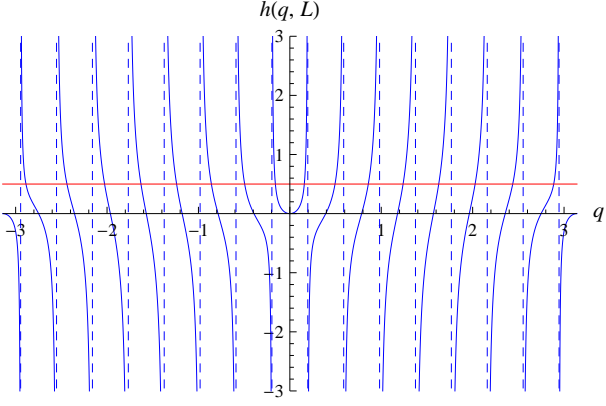}
\caption{(Color online) The function $h(q,L)$ [Eq.~\eq{DISPLANA}] for $L=16$. The vertical dashed lines represent the asymptotes of $h(q,L)$; the horizontal line stands for ${\cal K} = \half $.}
\label{HFIG}
\end{figure}

The quantization condition for $q$ now becomes
\beq
h(q,L)=\sin q  \tan\frac{qL}{2} = \Kappa \, . \label{DISPLANA}
\eeq
We plot the function $h(q,L)$ for $L=16$ in Fig.~\ref{HFIG}; the horizontal line depicts the value of the right-hand side of Eq.~\eq{DISPLANA} for $\Kappa =  \half$. The plot shows that the function $h(q,L)$ attains every real value $> 0$ precisely $L$ times when $q$ varies in the interval $(-\pi,\pi)$, once in each interval of the form $(\pi Q/L, \pi Q/L +\pi/L)$ for an integer $Q$.

Let us again for definiteness take $\Kappa>0$ and consider nonnegative roots $q\in[0,\pi)$ to Eq.~\eq{DISPLANA} (if $q$ is a root, then so is $-q$). Such roots for $\Kappa=0$ would be
\beq
p_Q = \frac{2\pi Q}{L},\quad Q=0,\ldots,L/2-1\,.
\label{PQEQ}
\eeq
We characterize the roots $q_Q$ for $\Kappa>0$ with the same ``quantum number'' $Q$ in such a way that $q_Q$ develops continuously from $p_Q$ when $\Kappa$ is continuously increased from zero. In the limit $L\rightarrow\infty$ the change in $\sin q$ is negligible for the change of $\Kappa$ from zero to its final value, so that the actual energy value satisfies $\omega_{c} (Q)= -\cos q_Q\simeq -\cos p_Q$, and therefore $\sin q_Q \simeq \sqrt{1-\cos^2 p_Q}$. Noting this, one immediately sees that to the leading nontrivial order in $1/L$, the $Q{\rm th}$ root of Eq.~\eq{DISPLANA} is
\beq
q_Q \simeq p_Q+\frac{2}{L} \arctan\frac{\Kappa}{\sqrt{1-\cos^2 p_Q}}\equiv
p_Q + \Delta_Q\,.
\label{APPRCR}
\eeq
The branch of the explicit $\arctan$ function is chosen so that the value lies between $-\half\pi$ and $\half\pi$, and the addition of $p_Q$ corresponds to the choice of the proper branch of the $\arctan$ function to solve Eq.~\eq{DISPLANA}.

After this organizational work, the roots of Eq.~\eq{DISPLANA} and the corresponding energies $\omega_{c}(Q)= -\cos q_Q$ are easy to find numerically. The results agree with those obtained by solving Eq.~\eq{ENEQ} numerically. Again, if the finite number of lattice sites is the issue, we recommend direct numerical computations.

The state vector in momentum representation is the inverse of the discrete Fourier transformation \eq{DISCFT}, or
\beq
A_c(\omega_{c},p) = \frac{1}{\sqrt{L}} \sum_{k=-L/2}^{L/2-1}e^{-ipk}  \alpha_k(\omega_{c})\,,
\eeq
where the momenta $p$, in fact, run over the values called $p_Q$ in Eq.~\eq{PQEQ}, albeit with $Q = -L/2 + 1, \ldots, L/2$. Inserting Eq.~\eq{ANSCON}, making use of Eqs.~\eq{CCOND} and \eq{DISPLANA}, and noting that $\sin(pL/2)=0$, we have
\beq
A_c(\omega_{c},p) = \frac{D \Kappa}{\sqrt{L}} \frac{1}{\cos p - \cos q}\,,
\label{NEWP}
\eeq
where $D$ is a so-far undetermined normalization constant. Given that $\cos q=-\omega_{c}$, this is in agreement with the previous expression for the state vector \eq{ONCEMORE}.

The problem in an attempt to take the limit $L\rightarrow\infty$ is the terms with $p\simeq q$, which make the expression~\eq{NEWP} singular. However, by virtue of Eq.~\eq{APPRCR}, we are now in a position to take the continuum limit differently than we did in the momentum representation. Following an old~\cite{RIC29,FAN35} but maybe not so generally known idea, let us consider Eq.~\eq{NEWP} in the vicinity of the singularity that occurs around a given $q$ value labeled $q_{Q_0}$, with $p >0$ and $q>0$ for definiteness.  We then have
\beq
A_c(\omega_{c},p_Q) \simeq -\frac{DK}{\sqrt{L}}\, \frac{1}{\sin p_{Q_0}(p_Q - p_{Q_0} - \Delta_{Q_0})}\,.
\eeq
In the limit $L \rightarrow \infty$, an arbitrary fixed small symmetric neighborhood $\Delta p$ of the momentum $p_{Q_0}$ contains a very large number of momentum modes $p_Q$. Taking a smooth function of momentum $G(p)$, we estimate
\bea
 &&\sum_{p_Q\in \Delta p} A_c(\omega_{c},p_Q) G(p_Q)\nonumber\\
&&\simeq - \frac{D\Kappa}{\sqrt{L}} \frac{2\pi}{L \sin p_{Q_0}} G(p_{Q_0})\sum_{Q=-\infty}^\infty\frac{1}{Q\!-\!Q_0\! -\! L \Delta_{Q_0}/(2 \pi) } \nonumber \\
 \commentout{ && =  -\frac{D\Kappa}{\sqrt{L}} \frac{2\pi}{L \sin p_{Q_0}} G(p_{Q_0})\!\!\!\sum_{Q=-\infty}^\infty \frac{1}{2} \bigg{[} \frac{1}{Q\!-\! L \Delta_{Q_0}/(2 \pi)} \! + \!\frac{1}{-Q- L \Delta_{Q_0}/(2 \pi)} \bigg{]} \nonumber \\}
  && = -\frac{DK}{\sqrt{L}} \frac{2\pi}{L \sin p_{Q_0}} G(p_{Q_0})\!\!\! \sum_{Q=-\infty}^\infty\left[\frac{ L \Delta_{Q_0}/(2 \pi)}{Q^2\!-\! (L \Delta_{Q_0}/(2 \pi))^2} \right]\!,\nonumber\\
\eea
from duplication of the original sum with the change of the summation index $Q\rightarrow-Q$.
However, the sum~\cite{FOOTNOTE1}
\beq
\sum_{k=-\infty}^\infty \frac{x}{k^2-x^2} = -\pi \cot[\pi x]
\label{COTSUM}
\eeq
and the definition of $\Delta_{Q_0}$ from Eq.~\eq{APPRCR} give
\nopagebreak[1]
\bea
&& \sum_{p_Q\in \Delta p} A_c(\omega_{c},p_Q) G(p_Q)
\nonumber \\ && \simeq \frac{D\Kappa}{\sqrt{L}} \,\frac{L}{2\pi} G(p_{Q_0}) \left[ \frac{1}{\sin p_{Q_0}} \frac{\pi\sqrt{1-\cos^2 p_{Q_0}}}{\Kappa} \right].
\eea

On the other hand, replacing the sum with momentum points spaced at the intervals $2\pi/L$ by an integral, we find
\bea
&& \frac{2\pi}{L} \sum_{p_Q \in \Delta p} A_c(\omega_{c},p_Q) G(p_Q) \simeq \int_{p\in\Delta p} A_c(\omega_{c},p) G(p) \nonumber \\ && \simeq \frac{D\Kappa}{\sqrt{L}} \left[ \frac{1}{\sin p_{Q_0}} \frac{\pi\sqrt{1-\cos^2 p_{Q_0}}}{\Kappa} \right]G(p_{Q_0}) \, .
\eea
Clearly, in the small range $\Delta p$, the function $A_c(\omega_c,p)$ behaves as
\bea
A_c(\omega_{c},p) &\simeq& \frac{D\Kappa}{\sqrt{L}}
\left[
\frac{1}{\sin p_{Q_0}}\frac{\pi\sqrt{1-\cos^2 p_{Q_0}}}{\Kappa}
\right]\delta(p-p_{Q_0})\nonumber\\
&\simeq& \frac{D\Kappa}{\sqrt{L}}\, \frac{\pi\sqrt{1-\omega_{c}^2}}{\Kappa}\, \delta(\cos p + \omega_{c})\,,
\eea
where we have noted in the last step that $\omega_{c}\simeq -\cos p_{Q_0}$. Other than in the small and symmetric neighborhood $\Delta p$, Eq.~\eq{NEWP} still applies. In the limit $L\rightarrow\infty$, the momentum representation state vector therefore behaves under integrals over $p$ as
\beq
A_c(\omega_{c},p)  =  \frac{D\Kappa}{\sqrt{L}} \Bigg{[}
\textrm{P}\frac{1}{\omega_{c}+\cos p} + \frac{\pi\sqrt{1-\omega_{c}^2}}{\Kappa}\, \delta(\omega_{c}+\cos p)
\Bigg{]}.\\
\label{CSTATE1}
\eeq

The result has the same functional form as before [Eq.~\eq{CSTATE}] so all that remains is to verify the normalization constant. To this end we first note that the condition for normalization of the state \eq{ANSCON}
\beq
\sum_k \alpha_k(\omega_{c})^2 = \half
\eeq
is cast in the form
\beq
D = \sqrt\frac{1-\omega_{c}^2}{L(\Kappa^2+1-\omega_{c}^2)-2 \Kappa \omega_{c}}
\eeq
by using Eqs.~\eq{CCOND} and \eq{DISPLANA}. In the limit $L\rightarrow\infty$, Eq.~\eq{CSTATE} immediately follows.

\section{Examples of Dimer  Detection}

\subsection{Momentum distribution}
In our first foray into the detection of the dimers we assume that after a preparation of a possibly large number of bound-state dimers, the lattice is removed and the atomic cloud expands ballistically. The positions of the atoms are then detected after some free-flight time. Ideally, this procedure converts the momentum distribution of the atoms into a position distribution so that a measurement of the position distribution amounts to a measurement of the momentum distribution. Our thought experiments closely mimic actual laboratory experiments~\cite{WinklerK441}.

There is a complication arising from the periodicity of the lattice that was also discussed in~Ref.~\cite{WinklerK441}. Namely, if the lattice is switched off instantaneously, the momentum distribution consists of periodic repetitions of the first Brillouin zone modulated by the momentum distribution of an atom in one individual lattice site, that is, the momentum distribution of the one-atom states associated with the annihilation operators $a_k$. In the experiments~\cite{WinklerK441} the lattice was turned off on a time scale such that the structure of the physics on a length scale below one lattice spacing was presumably removed adiabatically, while the site-to-site physics did not have time to adjust. The result is a momentum distribution confined to the first Brillouin zone. We analyze such distributions as well.

Now take an eigenstate of the center-of-mass motion~\eq{ANSATZ} with the total momentum $P$. A straightforward exercise then gives the momentum distribution,
\beq
{\cal M}(p) = \mele{\psi}{c^\dagger_pc_p}{\psi} = 4 \left| A_b \left( \omega_b , p - \half P \right) \right|^2 \, ,
\label{MOMDIST}
\eeq
with the $A$ function given in Eq.~\eq{BSTATE} for the bound state.

First consider a stationary center of mass, $P=0$; then we have
\beq
{\cal M}(p) \propto \frac{1}{(\omega_b + \cos p)^2}\,.
\eeq
For a repulsively bound state, $\omega_b > 1$ holds true, so that the momentum distribution is a maximum at the edges of the first Brillouin zone, when $\cos p = -1$, or $p = \pm \pi$.  For the bound state with attractive interactions, the maximum is at the center, $p = 0$, of the first Brillouin zone. These features were seen experimentally~\cite{WinklerK441}.

\begin{figure}
\includegraphics[width=8.0cm]{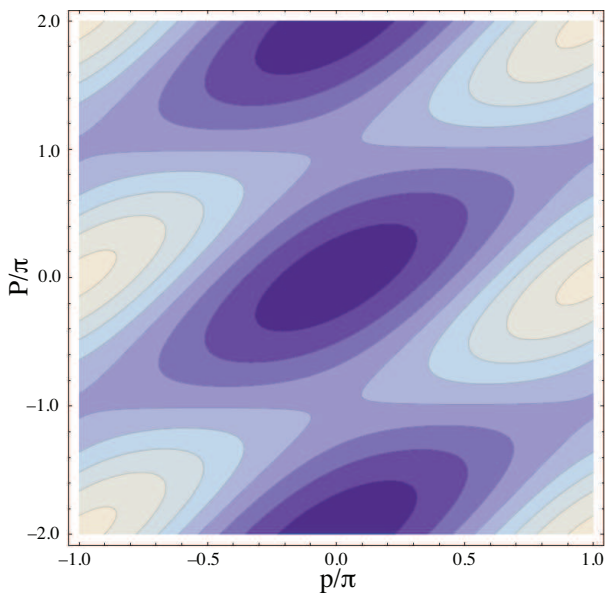}
\caption{(Color online) Contour plot of the momentum distribution in the bound state of the lattice dimer, $f(p,P)$ of Eq.~\eq{CPMD}, as a function of the momenta of the individual atoms $p$ and the center-of-mass momentum $P$, for $U/J=16$. Brighter shades represent larger values.}
\label{CPL}
\end{figure}

At first it might seem that the variation of Eq.~\eq{MOMDIST} with the center-of-mass momentum $P$ would simply be to slide around the momentum distribution of the atoms cyclically in the first Brillouin zone by $\half P$. This is not the case, since the energy of the bound state $\omega_b$ also depends implicitly on $P$. For convenience, we define a momentum distribution $f(p,P) = L {\cal M}(p)/4 \pi$ normalized so that $\int_{-\pi}^{\pi} \! f(p,P) \, dp = 1$ and have the explicit expression
\bea
&&\!\!\!\!\!\!\!\!f(p,P) = \frac{|{\cal K}(P)|^3}{2\pi\sqrt{1+{\cal K}(P)^2}
}\nonumber\\
&&\!\!\!\!\!\times \frac{1}{\left\{
\cos(p-\half P) +{\rm sgn}[{\cal K}(P)]\sqrt{1+{\cal K}(P)^{\!2}}
\right\}^2}\,,
\label{CPMD}
\eea
with
\beq
{\cal K}(P) = \frac{{\cal K}_0}{\cos\half P},\quad {\cal K}_0 = \frac{U}{2J}\,.
\eeq
We plot an example with ${\cal K}_0=8$ in Fig.~\ref{CPL}. This is a contour plot with $p$ as the horizontal axis, $P$ as the vertical axis, and brighter shades standing for larger values. The $\half P$ sliding of the distribution of the atomic momenta $p$ with the center-of-mass momentum $P$ is visible, but there is also a modulation so that, if anywhere, the momentum distribution of  the atoms is always peaked near the edges of the Brillouin zone. The peaks in the momentum distribution are narrowest for the center-of-mass momenta $P=-2\pi$, $0$, and $P = 2 \pi$; at $P = \pm \pi$, the momentum distribution is completely flat, $f(p,\pm\pi)= 1 / 2 \pi$.

A discussion of the momentum distribution as a function of the center-of-mass momentum based on numerical solutions of the lattice as a many-body system was also offered in Ref.~\cite{WinklerK441}. It brings up similar qualitative elements as our discussion, but it seems to us that the translation of the momentum distribution with the center-of-mass momentum is quoted  in Ref.~\cite{WinklerK441} as $P$, whereas we obtain $\half  P$.

\subsection{Size of the bound state}\label{PAIRCORRELATIONS}

Our next example on the detection of the dimer is a thought experiment in which there are precisely two atoms in the lattice, and the number of the atoms at each site is measured. This experiment is carried out over and over again, and the detection statistics is compiled. In our example we assume an absorbing detector that removes an atom from further consideration once it has been observed. Modeling after the well-known photon detection theory~\cite{GLA63,KEL64}, the joint probability for finding an atom at sites $k_1$ and $k_2$ in the energy eigenstate of the lattice dimer~\eq{TRUEEIGNSTATE} with a fixed center-of-mass momentum $P$ is
\bea
{\cal J}(k_1,k_2) & = & N \mele{\omega}{a^\dagger_{k_1} a^\dagger_{k_2}a_{k_2}a_{k_1}}{\omega} \nonumber \nopagebreak[1] \\  & = & \frac{4N}{L} |\alpha_{k_1-k_2}|^2 \, ,
\eea
where $N$ is a normalization constant. As befits translational invariance, the probability only depends on the distance between the lattice sites. The distribution of the distance is governed by the internal state of the molecule in position representation, $\alpha_k$ of Sec.~\ref{SOLPOSREP}.

We will not analyze the implications for a finite size lattice, but go directly to the limit of infinite lattice with $L \rightarrow \infty$. By virtue of Eq.~\eq{POSWF} the variance of the distance between the detected atoms in the bound state is
\beq
(\Delta k)^2 = \frac{\sum_k k^2 |\alpha_k|^2}{\sum_k|\alpha_k|^2}   = \frac{1}{2\Kappa^2}\,.
\eeq

Any eigenstate~\eq{EINSI} for a fixed center-of-mass momentum $P$ of the two bosons in the lattice is translation invariant. Taken individually, the detected atoms must be evenly distributed along the lattice, and there is no sign of the internal state of the dimer. The internal state is manifested in the correlations between the detected positions of the atoms; in the ground state the atoms are observed in pairs with the root mean square distance $1/(\sqrt{2}|\Kappa|)$ between them. Even if the interactions between the atoms are repulsive, the stronger the interactions, the more tightly the atoms are paired.

\subsection{Dissociation rate of the bound state}\label{DRBS}
In order to find the bound-state energy experimentally, some form of spectroscopy has to be carried out. Here we assume that the intensity of the lattice light is modulated periodically. The dissociation rate of a repulsively bound pair ($U>0$) as a function of the modulation frequency was studied experimentally in this way in Ref.~\cite{WinklerK441}.

Given that the tunneling amplitude is much more sensitive to the depth of the optical lattice than the atom-atom interactions,  in our model only the hopping matrix element $J$ becomes time dependent:
\beq
J \quad\rightarrow\quad J(1 + \Lambda \cos\nu t)\,,
\eeq
where $\Lambda J$ and $\nu$ are the amplitude and frequency of the modulation, respectively. The modulation adds a ``perturbation'' to the Hamiltonian. In lattice momentum representation, we write it as
\beq
\frac{H'}{\hbar} =\Lambda \cos\nu t \,\sum_q \omega_q\,c^\dagger_q c_q \, .
\eeq

The matrix element of this perturbation between any two states $i$ and $j$ of the form \eq{ANSATZ} turns out to be nonzero only if the states have the same center-of-mass momentum $P$; then we have
\beq
\mele{\psi_i}{\frac{H'}{\hbar}}{\psi_j} = -\Lambda\Omega_P M_{ij}\,\cos\nu t\,.
\eeq
It proves expedient to define dimensionless matrix elements, scaled to the frequency $\Omega_P$, as
\nopagebreak[1]
\bea
M_{ij} & = & 2 \sum_q \cos q\, A^*_i(q) A_j(q) \nonumber \\ & \simeq & \frac{L}{\pi} \int_{-\pi}^{\pi} dq\,\cos q \,A^*_i(q) A_j(q) \, .
\label{MELEINTEG}
\eea
Some results in what follows have maybe unexpected complications in the notation, which are needed to cover the possibility that $\Omega_P<0$.

The relevant $L\rightarrow\infty$ energy eigenstates are specified by Eqs.~\eq{BSTATE} and \eq{CSTATE} . The matrix elements are
\bea
&&\!\!\!\!\!\!M_{bb}  \simeq  -\frac{1}{\omega_b}\,, \\
&&\!\!\!\!\!\!M_{bc} = M_{cb}  \simeq   \left [\frac{2{\cal K}^3 \left(1-\omega_c^2\right )}{L\left (\omega_b^2 -\omega_c^2 \right)\omega_b}\right ]^{\frac{1}{2}}\,,\\
&&\!\!\!\!\!\!M_{cc'} \simeq -\frac{2 \pi\omega_c \sqrt{1-\omega_c^2}}{L}\,\delta(\omega_c-\omega_{c'})\nonumber\\&& + \frac{2 {\cal K}}{L}\sqrt\frac{(1-\omega_c)(1-\omega_c')}{({\cal K}^2+1-\omega_c)({\cal K}^2+1-\omega_c')}\,.
\label{CONMATELE}
\eea
The matrix elements $M_{cc'}$ may look severely singular, but they are not. Namely, recognizing $2\pi \sqrt{1-\omega_c^2}/L$ as the inverse of the density of the states $\omega_c$, the finite-lattice version of the delta function part is
\bea
-\frac{2 \pi\omega_c \sqrt{1-\omega_c^2}}{L}\,\delta(\omega_c-\omega_{c'})\sim -\omega_c \, \delta_{\omega_c,\omega_{c'}} \, .
\label{UDSTD}
\eea
This is a perfectly well behaved diagonal matrix element for the continuum states.

The form for the amplitudes $A$ in Eq.~\eq{CSTATE} was supposed to be good inside integrals only if the rest of the integrand is well behaved, which is clearly not the case in Eq.~\eq{MELEINTEG} when $i$ and $j$ both stand for continuum states. The matrix elements~\eq{CONMATELE}, however, transcend their derivation, with the understanding~\eq{UDSTD} that they agree with the matrix elements computed numerically in a large finite lattice.

Writing the state $|\omega\rangle$ as $|b\rangle$ for the bound state and as $|c\rangle$ for the quasicontinuum state, the Hamiltonian written in the eigenbasis of $H_0$ is
\bea
&&\frac{H}{\hbar\Omega_P}  =  \omega_b \dyad{b}{b} + \sum_{c} \omega_c \dyad{c}{c} \nonumber \\ && - \Lambda\cos\nu t \bigg{[} M_{bb} \dyad{b}{b} + \sum_{c} \left( M_{bc}\dyad{b}{c} + M_{cb} \dyad{c}{b} \right) \nonumber \\ && + \sum_{cc'}M_{cc'}\dyad{c}{c'} \bigg{]} \, .
\eea
We employ perturbation theory in the dimensionless parameter characterizing the modulation depth, $\Lambda$, to study the dissociation rate of the bound state to the quasicontinuum states. Here the problem of time-dependent perturbation theory is unusual in that there are diagonal transition matrix elements. We attempt to get past this obstacle with the assumption that the system starts in the bound state. The matrix elements $M_{bc}$ and $M_{cb}$ must then be involved. The corresponding terms in the Hamiltonian are already in themselves first order in the small parameter $\Lambda$. We therefore apply the Hamiltonian in the form
\bea
\frac{H}{\hbar\Omega_P} & = & \omega_b\dyad{b}{b}
+ \sum_{c} \omega_c\dyad{c}{c} \nonumber \\ && - \Lambda\cos\nu t \sum_{c} \left( M_{bc} \dyad{b}{c} + M_{cb}\dyad{c}{b} \right)
\eea
in the hope that we get a correct description to leading order in perturbation theory in the parameter $\Lambda$. This should be warranted if the coupling between the continuum states mediated by the matrix elements $M_{cc'}$ does not cause the exact solution of the time-dependent Schr\"odinger equation to become a nonanalytic function of the parameter $\Lambda$ at $\Lambda=0$. We assume so without further ado.

At this point the standard Golden Rule transition rate applies and gives the dissociation rate of the bound state in proper dimensional units as
\beq
\Gamma = \frac{\pi\Lambda^2 |\Omega_P||M_{bc}|^2}{2} \varrho(\Delta).
\eeq
Here the energy-conserving continuum state with the label $\omega_c\equiv\Delta$ depends on whether the bound state lies above ($-$) or below ($+$) the continuum,
\beq
\Delta = \frac{E_b}{\hbar\Omega_P} \mp \frac{\nu}{\Omega_P}\,.
\eeq
This is an analog of a parameter called detuning in laser spectroscopy, and would be controlled in practice by varying the modulation frequency $\nu$. Using the density of states given in Eq.~\eq{DENSITYOFSTATES}, we have
\beq
\Gamma = \frac{ \Lambda^2\,|\Omega_P||{\cal K}|^3 \sqrt{1-\Delta^2}}{2|\omega_b|(\omega_b^2-\Delta^2)}\,.
\label{LINESHAPE}
\eeq

The shape of the dissociation line, variation of the dissociation rate with the modulation frequency, changes from $(1-\Delta^2)^{-1/2}$ in the limit of weak interactions $|{\cal K}|\ll1$ to $(1-\Delta^2)^{1/2}$ in the limit $|{\cal K}|\gg1$. In dimensional units the width of the spectrum is $2|\Omega_P|$.
The total strength of the transition is characterized by
\beq
\int_{-1}^{1}d\Delta\, \frac{\Gamma(\Delta)}{|\Omega_P|\Lambda^2} = \frac{\pi {|\cal K}|^3}{2(1+{\cal K}^2)(|{\cal K}|+\sqrt{1+{\cal K}^2})}\,,
\eeq
which tends to $\pi|{\cal K}|^3/2$ as $|{\cal K}|\ll1$ and to $\pi/4$ as $|{\cal K}|\gg1$.

The experimental results~\cite{WinklerK441} are mostly in the limit $\Kappa \gg 1$. While there is no indication in Ref.~\cite{WinklerK441} whether perturbation theory in the modulation amplitude should apply or not, our results qualitatively explain the published line shape.
\commentout{Our results qualitatively explain the published line shapes. On the other hand,  there is no indication in Ref.~\cite{WinklerK441} whether perturbation theory in the modulation amplitude should apply or not, so that it need not come as a surprise that a quantitative comparison between our development and the experiments runs into problems. For instance, the width of the dissociation lineshape in Fig. 4~(a) of Ref.~\cite{WinklerK441} is about twice the width of the continuum band for the stated parameters.}

\section{Concluding Remarks}

We have demonstrated how various molecular physics, condensed-matter physics, and many-body physics aspects come together in a description of a dimer of two bosonic atoms in an optical lattice.  Aside from a surprising number of mathematical complications that we have sorted out both explicitly and behind the scenes, the main technical issue here is the old~\cite{RIC29,FAN35} quasicontinuum problem: what to do with a system that has a dense set of energy eigenstates? We have described the stationary states of the dimer and found analytical expressions for the stationary states in the limit of an infinitely long lattice. Once the groundwork is laid, applications are straightforward. As an example, we have briefly discussed three ways to detect a bound dimer.

We have analyzed the one-dimensional Bose-Hubbard model as a specific example. Nevertheless, we can think up, and several authors~\cite{NYG08FES, GRU07, PII08DIF, NguenangJ80} have thought up, a number of similar problems. Ultimately, what we hope to have achieved here is a template, a methodology, that applies to all sorts of dimer problems in lattices and will possibly contribute to future studies of aggregates of atoms in a lattice with more than two atoms as well.

\section*{Acknowledgments}
This work is supported in part by NSF Grant No.\ PHY-0651745.

\appendix
\section{Operator Algebra Example}
The second-quantized notation automatically takes care of the symmetries with respect to particle exchange, but we may run into possibly long products of creation and annihilation operators. In some subfields of physics the methods for dealing with them are standard fare, but for completeness we demonstrate a technique employed frequently in this article by deriving Eq.~\eq{ACTINTER}. First, given Eq.~\eq{ANSATZOLD}, we have from the left-hand side of Eq.~\eq{ACTINTER}
\begin{widetext}
\beq
\sum_{q_1,q_2,q_3,q_4}
\delta_{[q_1+q_2,q_3+q_4]}\,
c^\dagger_{q_1}c^\dagger_{q_2}c_{q_3}c_{q_4} \ket\psi =
\sum_{q_1,q_2,q_3,q_4,p_1,p_2}  A(p_1,p_2)\,
\delta_{[q_1+q_2,q_3+q_4]}\,
c^\dagger_{q_1}c^\dagger_{q_2}c_{q_3}c_{q_4} c^\dagger_{p_1}c^\dagger_{p_2} \ket0 \, .
\label{AP0}
\eeq
The general idea is to put the products of the operators into normal order, annihilation operators to the right of creation operators, by using the commutators (or anticommutators) of the operators. Here  $c_p c^\dagger_q = c^\dagger_q c_p + \delta_{[p,q]}$ holds for arbitrary $p$ and $q$. We therefore have the following chain of manipulations:
\bea
c^\dagger_{q_1}c^\dagger_{q_2}c_{q_3}c_{q_4} c^\dagger_{p_1}c^\dagger_{p_2} \ket0 &=&
c^\dagger_{q_1}c^\dagger_{q_2}(c_{q_3}c^\dagger_{p_1}c_{q_4} c^\dagger_{p_2}
+\delta_{[q_4,p_1]}c_{q_3}c^\dagger_{p_2})\ket0
=c^\dagger_{q_1}c^\dagger_{q_2}(\delta_{[q_4,p_2]}c_{q_3}c^\dagger_{p_1}
+\delta_{[q_4,p_1]}\delta_{[q_3,p_2]})\ket0\nonumber\\
&=& c^\dagger_{q_1}c^\dagger_{q_2}(\delta_{[q_4,p_2]}\delta_{[q_3,p_1]}
+\delta_{[q_4,p_1]}\delta_{[q_3,p_2]})\ket0;
\label{AP1}
\eea
once an annihilation operator has been moved to operate on the vacuum $\ket0$, the corresponding term vanishes. Sorting out the Kronecker $\delta$'s and using the symmetry $A(p_1,p_2)=A(p_2,p_1)$,~\eq{AP0} and~\eq{AP1} immediately combine to give the right-hand side of Eq.~\eq{ACTINTER}.
\end{widetext}

\bibliography{MasterJJPaperV110}

\begin{thebibliography}{25}
\bibitem [{FOO()}]{FOOTNOTE1}%
  \BibitemOpen
  \href@noop {} {}\bibinfo {note} {{\rm This sum is known to v. 7.0 of
  MATHEMATICA}. It may be calculated analytically by converting the sum to a
  contour integral with the aid of the function $(e^{2\pi i z}-1)^{-1}$, which
  has a simple pole at all integer values of $z$}\BibitemShut {NoStop}%
\end{thebibliography}

\end{document}